\documentclass[12pt]{article}
\usepackage{amsmath}
\usepackage{amssymb}
\usepackage{amsfonts}

\numberwithin{equation}{section}
\newcommand{\chh}{{\cal H}}
\newcommand{\crr}{{\cal R}}
\newcommand{\cii}{{\cal I}}
\newcommand{\cxx}{{\cal X}}
\newcommand{\cyy}{{\cal Y}}
\newcommand{\field}[1]{\mathbb{#1}}
\newcommand{\Z}{\field{Z}}

\title{
Superintegrability of the Tremblay-Turbiner-Winternitz quantum Hamiltonians on a plane for odd $k$}
\author{C Quesne\\ 
{\small Physique Nucl\'eaire Th\'eorique et Physique Math\'ematique,  Universit\'e Libre de Bruxelles,} \\ 
{\small Campus de la Plaine CP229, Boulevard~du Triomphe, B-1050 Brussels, Belgium}}
\date{ }
\begin{document}
\baselineskip=22pt plus 1pt minus 1pt
%%%%%%%%%%%%%%%%%%%%%%%%%%%%%%%%%%%%%%%%%%%%%%%%%%%%%%%%%%
\maketitle

\begin{abstract} 
In a recent FTC by Tremblay {\sl et al} (2009 {\sl J.\ Phys.\ A: Math.\ Theor.} {\bf 42} 205206), it has been conjectured that for any integer value of $k$, some novel exactly solvable and integrable quantum Hamiltonian $H_k$ on a plane is superintegrable and that the additional integral of motion is a $2k$th-order differential operator $Y_{2k}$. Here we demonstrate the conjecture for the infinite family of Hamiltonians $H_k$ with odd $k \ge 3$, whose first member corresponds to  the three-body Calogero-Marchioro-Wolfes model after elimination of the centre-of-mass motion. Our approach is based on the construction of some $D_{2k}$-extended and invariant Hamiltonian $\chh_k$, which can be interpreted as a modified boson oscillator Hamiltonian. The latter is then shown to possess a $D_{2k}$-invariant integral of motion $\cyy_{2k}$, from which $Y_{2k}$ can be obtained by projection in the $D_{2k}$ identity representation space.

\end{abstract}


%
%========================================================================
%
\newpage
\section{Introduction}

In a recent work, Tremblay, Turbiner and Winternitz (TTW) have introduced an infinite family of exactly solvable quantum Hamiltonians on a plane
\begin{equation}
  H_k = - \partial_r^2 - \frac{1}{r} \partial_r - \frac{1}{r^2} \partial_{\varphi}^2 + \omega^2 r^2 +
  \frac{k^2}{r^2} (\alpha \sec^2 k\varphi + \beta \csc^2 k\varphi),  \label{eq:H-k}
\end{equation}
defined in terms of polar coordinates $(r, \varphi)$ and of an arbitrary real number $k$ \cite{tremblay09}. In (\ref{eq:H-k}), the configuration space is given by the sector $0 \le r < \infty$, $0 \le \varphi \le \pi/(2k)$, and $\omega$, $\alpha$, $\beta$ are three parameters such that $\omega > 0$, $\alpha = a(a-1) > - 1/(4k^2)$, $\beta = b(b-1) > - 1/(4k^2)$. The Hamiltonian $H_k$ has been shown to be integrable with a second-order integral of motion
\begin{equation}
  X_k = - \partial_{\varphi}^2 + k^2 (\alpha \sec^2 k\varphi + \beta \csc^2 k\varphi),  \label{eq:X-k}
\end{equation}
responsible for the separation of variables of the corresponding Schr\"odinger equation in polar coordinates.\par
%
%-----------------------------------------------------------------------------------------------------
%
In the same work, it has also been conjectured that for all integer values of $k$, the Hamiltonian (\ref{eq:H-k}) is superintegrable and that the additional functionally-independent integral of motion is a $2k$th-order differential operator $Y_{2k}$. This conjecture has been proved by a direct calculation of $Y_{2k}$ for $k=1$, 2, 3 and 4, as well as for $k=1$ to 6 and 8 in the special case where either $\alpha$ or $\beta$ vanishes. Actually, for $k=1$, 2 and 3, the superintegrability of $H_k$ was already known since the Hamiltonian then reduces to that of the Smorodinsky-Winternitz system ($k=1$) \cite{fris, winternitz}, of the rational $BC_2$ model ($k=2$) \cite{olsha} or of the Calogero-Marchioro-Wolfes (CMW) model ($k=3$) \cite{calogero, wolfes}.\par
%
%-----------------------------------------------------------------------------------------------------------------------
%  
More recently, the credibility of the conjecture has been reinforced by showing that for the corresponding classical systems all bounded trajectories are closed and that the motion is periodic for all integer and rational values of $k$ \cite{tremblay10}. Such classical systems have then been proved to be superintegrable \cite{kalnins09a} and generalizable to higher dimensions \cite{kalnins09b}.\par
%
%---------------------------------------------------------------------------------------------------------------
%
Since it is clear that the proof of the conjecture for $k > 4$ requires a different approach from that based on a brute-force calculation of $Y_{2k}$, it is worth investigating the use of some differential-difference operators or covariant derivatives, known in the mathematical literature as Dunkl operators \cite{dunkl}. Such a formalism, based on the introduction of exchange operators, has indeed been successful in proving the superintegrability of the $N$-body Calogero model \cite{poly, brink}, which for $N=3$ is related to $H_3$ with $\beta = 0$, as well as that of the CMW model \cite{cq95}, for which some results in polar coordinates are also available \cite{khare}.\par
%
%-----------------------------------------------------------------------------------------------------------------
%
In a first step along these lines, we have generalized the exchange operators known for the CMW model to the whole set of Hamiltonians $H_k$ with integer $k$ \cite{cq09}. For this purpose, we have realized the elements of the dihedral group $D_{2k}$ as operators on the plane $(r, \varphi)$ and employed them to define some differential-difference operators $D_r$ and $D_{\varphi}$, where the latter assumes a different form according to whether $k$ is odd or even. We have then constructed $D_{2k}$-extended and invariant Hamiltonians $\chh_k$, from which the starting Hamiltonians $H_k$ can be retrieved by projection in the $D_{2k}$ identity representation space. Furthermore, we have shown that $X_k$ can be obtained by the same procedure from the operator $- D_{\varphi}^2$, commuting with $\chh_k$.\par
%
%-----------------------------------------------------------------------------------------------------------------
% 
The purpose of the present communication is to build on these advances in order to provide a rigorous proof of the above-mentioned conjecture for any odd $k$ value. More specifically, we plan to demonstrate the existence of a second integral of motion $Y_{2k}$ of $H_k$, obtainable by projection from a corresponding operator $\cyy_{2k}$ for the extended Hamiltonian $\chh_k$, whenever $k$ is odd and greater than or equal to three.\par
%
%===============================================================
%
\section{\boldmath Exchange operator formalism for $H_k$ with odd $k$}

In \cite{cq09}, it has been shown that for any integer $k$ value, the $4k$ elements of the dihedral group $D_{2k}$ can be realized as $\crr^i$ and $\crr^i \cii$, $i=0$, 1, \ldots, $2k-1$, where $\crr = \exp\left(\frac{1}{k} \pi \partial_{\varphi}\right)$ is the rotation operator through angle $\pi/k$ in the plane $(r, \varphi)$ while $\cii = \exp({\rm i} \pi \varphi \partial_{\varphi})$\footnote{In the definition of $\cii$ and only there, $\rm i$ stands for the square root of $-1$.} changes $\varphi$ into $-\varphi$. Such operators indeed satisfy the defining relations of (a unitary representation of) $D_{2k}$,
\begin{equation}
  \crr^{2k} = \cii^2 = 1, \qquad \cii \crr = \crr^{2k-1} \cii, \qquad \crr^{\dagger} = \crr^{2k-1}, \qquad 
  \cii^{\dagger} = \cii.  \label{eq:D-2k}  
\end{equation}
\par
%
%------------------------------------------------------------------------------------------------------------
%
The partial derivatives $\partial_r$ and $\partial_{\varphi}$ can then be extended into differential-difference operators $D_r$ and $D_{\varphi}$, defined for any odd $k$ by
\begin{equation}
\begin{split}
  D_r &= \partial_r - \frac{1}{r} (a\crr + b) \left(\sum_{i=0}^{k-1} \crr^{2i}\right) \cii, \\
  D_{\varphi} &= \partial_{\varphi} + a \sum_{i=0}^{k-1} \tan\left(\varphi + \frac{i\pi}{k}\right) \crr^{k+2i} \cii
        - b \sum_{i=0}^{k-1} \cot\left(\varphi + \frac{i\pi}{k}\right) \crr^{2i} \cii,
\end{split}  \label{eq:D}
\end{equation}
where $a$ and $b$ are related to the potential strengths $\alpha$ and $\beta$ in $H_k$. From (\ref{eq:H-k}), it follows that $D_r$ and $D_{\varphi}$ fulfil the following Hermiticity, exchange and commutation relations
\begin{equation}
  D_r^{\dagger} = - D_r - \frac{1}{r} \left[1 + 2(a\crr + b) \left(\sum_{i=0}^{k-1} \crr^{2i} \right) \cii\right], 
  \qquad D_{\varphi}^{\dagger} = - D_{\varphi},  \label{eq:D-hermite}
\end{equation}
\begin{equation*}
  \crr D_r = D_r \crr, \qquad \cii D_r = D_r \cii, \qquad \crr D_{\varphi} = D_{\varphi} \crr, \qquad 
  \cii D_{\varphi} = - D_{\varphi} \cii,
\end{equation*}
\begin{equation}
  [D_r, D_{\varphi}] = - \frac{2}{r} (a\crr + b) \left(\sum_{i=0}^{k-1} \crr^{2i} \right) \cii D_{\varphi}, 
  \label{eq:D-com} 
\end{equation}
respectively. It is worth noting that the first relation in (\ref{eq:D-hermite}) as well as equation (\ref{eq:D-com}) significantly differ from the corresponding results for $\partial_r$ and $\partial_{\varphi}$, namely $\partial_r^{\dagger} = - \partial_r - \frac{1}{r}$ and $[\partial_r, \partial_{\varphi}] = 0$.\par
%
%-------------------------------------------------------------------------------------------------------------------
%
{}From $D_r$ and $D_{\varphi}$, one can build a $D_{2k}$-extended Hamiltonian
\begin{equation}
  \chh_k = - D_r^2 - \frac{1}{r} \left[1 + 2 (a\crr + b) \left(\sum_{i=0}^{k-1} \crr^{2i}\right) \cii\right] D_r
  - \frac{1}{r^2} D_{\varphi}^2 + \omega^2 r^2,  \label{eq:ext-H-k}
\end{equation}
which can also be rewritten as
\begin{equation*}
  \chh_k = - \partial_r^2 - \frac{1}{r} \partial_r - \frac{1}{r^2} \left[D_{\varphi}^2 - k (a^2 + b^2 + 2ab \crr)
  \sum_{i=0}^{k-1} \crr^{2i}\right] + \omega^2 r^2
\end{equation*}
with
\begin{equation*}
\begin{split}
  D_{\varphi}^2 &= \partial_{\varphi}^2 - \sum_{i=0}^{k-1} \sec^2 \left(\varphi + \frac{i\pi}{k}\right) a (a -
       \crr^{k+2i} \cii) - \sum_{i=0}^{k-1} \csc^2 \left(\varphi + \frac{i\pi}{k}\right) b (b - \crr^{2i} \cii) \\
  & \quad + k (a^2 + b^2 + 2ab \crr) \sum_{i=0}^{k-1} \crr^{2i}.   
\end{split}
\end{equation*}
Such a $\chh_k$ is clearly invariant under $D_{2k}$ and, due to some simple trigonometric identites, gives rise to $H_k$ after projection in the $D_{2k}$ identity representation space, {\sl i.e.}, after replacing $\crr$ and $\cii$ by 1.\par
%
%------------------------------------------------------------------------------------------------------------------
%
{}Furthermore, $\cxx_k = - D_{\varphi}^2$ is a $D_{2k}$ invariant, commuting with $\chh_k$, and whose projection in the same space is given by $X_k - k^2 (a+b)^2$, with $X_k$ defined in equation (\ref{eq:X-k}).\par
%
%====================================================================
%
\section{\boldmath Modified boson operators for odd $k \ge 3$}

To go further, it is worth observing that, on one hand, the starting Hamiltonian $H_k$ has an oscillator spectrum \cite{tremblay09} and, on the other hand, the extended Hamiltonian $\chh_k$, defined in (\ref{eq:ext-H-k}), has (apart from some additional exchange operators) a form reminiscent of that of an oscillator Hamiltonian in polar coordinates, $h = - \partial_r^2 - \frac{1}{r} \partial_r - \frac{1}{r^2} \partial_{\varphi}^2 + \omega^2 r^2$. Since the latter can also be written as $h = 2\omega (h_a + h_b)$, $h_a = \frac{1}{2} \{a^{\dagger}, a\}$, $h_b = \frac{1}{2} \{b^{\dagger}, b\}$, in terms of two sets of boson creation and annihilation operators $a^{\dagger}$, $a$ and $b^{\dagger}$, $b$, it seems appropriate to introduce for $\chh_k$ two sets of modified boson operators
\begin{equation}
\begin{split}
  A &= \frac{1}{\sqrt{2\omega}} \left[\cos \varphi (\omega r + D_r) - \frac{1}{r} \sin \varphi D_{\varphi}\right],
      \\
  A^{\dagger} &= \frac{1}{\sqrt{2\omega}} \left[\cos \varphi (\omega r - D_r) + \frac{1}{r} \sin \varphi 
      D_{\varphi}\right],
\end{split}  \label{eq:A}
\end{equation}
and
\begin{equation}
\begin{split}  
  B &= \frac{1}{\sqrt{2\omega}} \left[\sin \varphi (\omega r + D_r) + \frac{1}{r} \cos \varphi D_{\varphi}\right],
      \\
  B^{\dagger} &= \frac{1}{\sqrt{2\omega}} \left[\sin \varphi (\omega r - D_r) - \frac{1}{r} \cos \varphi 
      D_{\varphi}\right],  
\end{split}  \label{eq:B}
\end{equation}
by simply substituting $D_r$ and $D_{\varphi}$ for $\partial_r$ and $\partial_{\varphi}$ in the definition of $a$, $a^{\dagger}$, $b$, $b^{\dagger}$ in polar coordinates.\par
%
%----------------------------------------------------------------------------------------------------------------
%
At this stage, it should be noted that the fact that $A^{\dagger}$ and $B^{\dagger}$ are the Hermitian conjugates of $A$ and $B$, respectively, is not so obvious as for $a^{\dagger}$ and $b^{\dagger}$ with respect to $a$ and $b$. To prove such a property, one has actually to use equations (\ref{eq:D-2k}), (\ref{eq:D-hermite}) and the definition of $\crr$ and $\cii$, as well as the commutation relations
\begin{equation}
  [D_r, r] = 1, \qquad [D_{\varphi}, r] = 0,  \label{eq:D-r}
\end{equation}
\begin{equation}
\begin{split}
  [D_r, \cos \varphi] &= \frac{1}{r} \biggl[2a \sum_{i=0}^{k-1} \cos \left(\varphi + \frac{i\pi}{k}\right)
        \cos \frac{i\pi}{k} \crr^{k+2i} \cii \\
  & \quad + 2b \sum_{i=0}^{k-1} \sin \left(\varphi + \frac{i\pi}{k} \right) \sin \frac{i\pi}{k} \crr^{2i} \cii \biggr],  
\end{split}  \label{eq:D-phi-1}
\end{equation}
\begin{equation}
\begin{split}
  [D_r, \sin \varphi] &= \frac{1}{r} \biggl[- 2a \sum_{i=0}^{k-1} \cos \left(\varphi + \frac{i\pi}{k}\right)
       \sin \frac{i\pi}{k} \crr^{k+2i} \cii \\ 
  & \quad + 2b \sum_{i=0}^{k-1} \sin \left(\varphi + \frac{i\pi}{k} \right) \cos  \frac{i\pi}{k} \crr^{2i} \cii \biggr], 
\end{split} 
\end{equation}
\begin{equation}
\begin{split}
  [D_{\varphi}, \cos \varphi] &= - \sin \varphi - 2a \sum_{i=0}^{k-1} \sin \left(\varphi + \frac{i\pi}{k}\right)
       \cos \frac{i\pi}{k} \crr^{k+2i} \cii \\
  & \quad + 2b \sum_{i=0}^{k-1} \cos \left(\varphi + \frac{i\pi}{k} \right) \sin \frac{i\pi}{k} \crr^{2i} \cii,  
\end{split}  
\end{equation}
\begin{equation}
\begin{split}
  [D_{\varphi}, \sin \varphi] &= \cos \varphi + 2a \sum_{i=0}^{k-1} \sin \left(\varphi + \frac{i\pi}{k}\right)
       \sin \frac{i\pi}{k} \crr^{k+2i} \cii \\
  & \quad + 2b \sum_{i=0}^{k-1} \cos \left(\varphi + \frac{i\pi}{k} \right) \cos \frac{i\pi}{k} \crr^{2i} \cii,  
\end{split}  \label{eq:D-phi-4}  
\end{equation}
which can be easily established from (\ref{eq:D-2k}) and (\ref{eq:D}).\par
%
%----------------------------------------------------------------------------------------------------------------------
%
It turns out, however, that the two sets of modified boson operators (\ref{eq:A}) and (\ref{eq:B}) do not have simple transformation properties under $D_{2k}$ whenever $k > 1$. To cope with this drawback, it is advantageous to replace $A$, $A^{\dagger}$ by $k$ dependent pairs of modified boson operators $A_i$, $A_i^{\dagger}$, $i=0$, 1, \ldots,~$k-1$, defined by
\begin{equation}
\begin{split}
  A_i &= \frac{1}{\sqrt{2\omega}} \left[\cos \left(\varphi + \frac{i\pi}{k}\right) (\omega r + D_r) - \frac{1}{r}
       \sin\left(\varphi + \frac{i\pi}{k}\right) D_{\varphi}\right], \\
  A_i^{\dagger} &= \frac{1}{\sqrt{2\omega}} \left[\cos \left(\varphi + \frac{i\pi}{k}\right) (\omega r - D_r) 
       + \frac{1}{r} \sin\left(\varphi + \frac{i\pi}{k}\right) D_{\varphi}\right],       
\end{split}  \label{eq:A-i}
\end{equation}
and such that $A_0 = A$ and $A_0^{\dagger} = A^{\dagger}$. In the same way, $B$, $B^{\dagger}$ may be replaced by $B_i$, $B_i^{\dagger}$, $i=0$, 1, \ldots,~$k-1$, obtained from (\ref{eq:B}) by substituting $\varphi + \frac{i\pi}{k}$ for $\varphi$ (hence $B_0 = B$, $B_0^{\dagger} = B^{\dagger}$). For most purposes, though, it will be enough to only consider $A_i$, $A_i^{\dagger}$, since $B_i$, $B_i^{\dagger}$ can be expressed in terms of them as\footnote{Up to equation (\ref{eq:A-i}), all mentioned results are valid for any odd $k$ value, including $k=1$. This is not the case for equation (\ref{eq:B-i}), nor for some formulas to be given in the sequel, which require $k \ge 3$. From now one, we therefore restrict ourselves to such values.} 
\begin{equation}
  B_i = \frac{2}{k} \sum_{j=0}^{k-1} A_j \sin \frac{(i-j)\pi}{k}, \qquad B_i^{\dagger} = \frac{2}{k} 
  \sum_{j=0}^{k-1} A_j^{\dagger} \sin \frac{(i-j)\pi}{k}.  \label{eq:B-i} 
\end{equation}
\par
%
%------------------------------------------------------------------------------------------------------------
%
Under $D_{2k}$, the operators $A_i$, $i=0$, 1, \ldots,~$k-1$, transform among themselves (and similarly for their Hermitian conjugates). We indeed obtain
\begin{equation}
\begin{split}
  \crr A_i \crr^{-1} &= A_{i+1}, \qquad i = 0, 1, \ldots, k-2, \qquad \crr A_{k-1} \crr^{-1} = - A_0, \\
  \cii A_0 \cii^{-1} & = A_0, \qquad \cii A_i \cii^{-1} = - A_{k-i}, \qquad i = 1, 2, \ldots, k-1. 
\end{split}  \label{eq:D-A}
\end{equation}
These relations can actually be simplified by extending the definition of $A_i$ (and $A_i^{\dagger}$) to all values $i \in \Z$,
\begin{equation}
  A_{\lambda k + i} = (-1)^{\lambda} A_i, \qquad i = 0, 1, \ldots, k-1, \qquad \lambda \in \Z.  \label{eq:A-Z}
\end{equation}
We then get
\begin{equation}
  \crr A_i \crr^{-1} = A_{i+1}, \qquad \cii A_i \cii^{-1} = - A_{k-i}, \qquad i \in \Z.  \label{eq:D-A-Z} 
\end{equation}
\par
%
%------------------------------------------------------------------------------------------------------------
%
Among the $k$ operators $A_i$, $i=0$, 1, \ldots,~$k-1$, there are only two linearly independent  ones ({\sl e.g.}, $A_0 = A$ and $B_0 = B$). The $A_i$'s therefore satisfy $k-2$ independent linear relations. We quote here some of them,
\begin{align}
  & \sum_{i=0}^{k-1} (-1)^i A_i = 0, \qquad \sum_{i=0}^{k-1} A_i  \cos \frac{i\pi}{k} = \frac{k}{2} A_0,
      \label{eq:A-rel-1} \\
  & A_i \cos \frac{i\pi}{k} = \frac{1}{2} (A_0 + A_{2i}), \qquad B_i \sin \frac{i\pi}{k} = \frac{1}{2} (A_0 - 
      A_{2i}), \qquad i\in \Z,  \label{eq:A-rel-2}  
\end{align}
which are easily derived through elementary trigonometry and will subsequently play an important role.\par
%
%---------------------------------------------------------------------------------------------------------------
%
The commutation relations of $A_i$, $A_i^{\dagger}$, $i=0$, 1, \ldots,~$k-1$, can be determined from equations (\ref{eq:D-com}) and (\ref{eq:D-r}), as well as from the relations obtained from (\ref{eq:D-phi-1}) -- (\ref{eq:D-phi-4}) by acting with $\crr^i$ on both sides. After a rather lengthy, but straightforward calculation, we get the simple results
\begin{equation}
  [A_i, A_j] = [A_i^{\dagger}, A_j^{\dagger}] = 0, \label{eq:A-com-1}  
\end{equation}
\begin{equation}
  \begin{split}
    [A_i, A_j^{\dagger}] &= [A_j, A_i^{\dagger}] = \cos \frac{(j-i)\pi}{k} + 2a \sum_l \cos \frac{(l-i)\pi}{k}
      \cos \frac{(l-j)\pi}{k} \crr^{k+2l} \cii  \\
    & \quad + 2b \sum_l \sin \frac{(l-i)\pi}{k} \sin \frac{(l-j)\pi}{k} \crr^{2l} \cii,  \label{eq:A-com-2}  
  \end{split}  
\end{equation}
for $i, j=0$, 1, \ldots,~$k-1$. Note that from now on, all summations over $l$ run from 0 to $k-1$.\par
%
%==================================================================
%
\section{\boldmath Modified boson oscillator Hamiltonians and superintegrability of $H_k$ for odd $k\ge 3$}

The modified boson operators $A_i$, $A_i^{\dagger}$, $i=0$, 1, \ldots,~$k-1$, can be used to define some modified oscillator Hamiltonians as
\begin{equation*}
  H_i = \tfrac{1}{2} \{A_i^{\dagger}, A_i\}, \qquad i = 0, 1, \ldots, k-1.
\end{equation*}
From (\ref{eq:D-A}), it follows that
\begin{equation}
\begin{split}
 \crr H_i \crr^{-1} &= H_{i+1}, \qquad i = 0, 1, \ldots, k-2, \qquad \crr H_{k-1} \crr^{-1} = H_0,  \\
 \cii H_0 \cii^{-1} &= H_0, \qquad \cii H_i \cii^{-1} = H_{k-i}, \qquad i = 1, 2, \ldots, k-1, 
\end{split}  \label{eq:D-H}
\end{equation}
so that the $H_i$'s transform among themselves under $D_{2k}$.\par
%
%--------------------------------------------------------------------------------------------------------------
%
The explicit expression of $H_i$ in terms of $r$, $\varphi$, $D_r$ and $D_{\varphi}$ can be easily found by first considering $i=0$ and then acting with $\crr^i$ on the result. It is given by
\begin{equation*}
\begin{split}
  2\omega H_i &= - \cos^2 \biggl(\varphi + \frac{i\pi}{k}\biggr) D_r^2 + \frac{1}{r} \sin \biggl(\varphi + 
      \frac{i\pi}{k}\biggr) \cos \biggl(\varphi + \frac{i\pi}{k}\biggr) (D_r D_{\varphi} + D_{\varphi} D_r) \\
  & \quad - \frac{1}{r^2} \sin^2 \biggl(\varphi + \frac{i\pi}{k}\biggr) D_{\varphi}^2 \\
  & \quad - \frac{1}{r} \biggl[\sin^2 \biggl(\varphi + \frac{i\pi}{k}\biggr) + 2a \sum_l \cos^2 \frac{(l-i)\pi}{k}
      \crr^{k+2l} \cii + 2b \sum_l \sin^2 \frac{(l-i)\pi}{k} \crr^{2l} \cii\biggr] D_r \\
  & \quad + \frac{1}{r^2} \biggl[- 2 \sin \biggl(\varphi + \frac{i\pi}{k}\biggr) \cos \biggl(\varphi + \frac{i\pi}{k}
      \biggr) - 2a \sum_l \sin \frac{(l-i)\pi}{k} \cos \frac{(l-i)\pi}{k} \crr^{k+2l} \cii \\
  & \quad + 2b \sum_l \sin \frac{(l-i)\pi}{k} \cos \frac{(l-i)\pi}{k} \crr^{2l} \cii\biggr] D_{\varphi} + \omega^2
      r^2 \cos^2 \biggl(\varphi + \frac{i\pi}{k}\biggr).   
\end{split}
\end{equation*}
\par
%
%-----------------------------------------------------------------------------------------------------------
%
Elementary trigonometry now shows that the $D_{2k}$-extended Hamiltonian $\chh_k$ of equation (\ref{eq:ext-H-k}) is directly connected with the $H_i$'s through the relation
\begin{equation}
  2\omega \sum_{i=0}^{k-1} H_i = \frac{k}{2} \chh_k.  \label{eq:H-sum}
\end{equation}
Hence, it may be regarded as a modified boson oscillator Hamiltonian.\par
%
%----------------------------------------------------------------------------------------------------------------
%
Next we may ask ourselves what are the commutation relations of the $H_i$'s. As explained below, for our purpose it is actually enough to determine the commutator of $H_0$ with $H_i$, $i=1$, 2, \ldots,~$k-1$. On using the convention (\ref{eq:A-Z}) for simplicity's sake, as well as equations (\ref{eq:D-A-Z}), (\ref{eq:A-com-1}) and (\ref{eq:A-com-2}), we get after some calculations
\begin{equation}
\begin{split}
  &[H_0, H_i] = (A_0^{\dagger} A_i - A_i^{\dagger} A_0) \cos \frac{i\pi}{k} + 2a \sum_l \biggl[- 
     (A_0^{\dagger} A_{2k+2l-i} + A_i^{\dagger} A_{k+2l}) \cos \frac{l\pi}{k} \cos \frac{(l-i)\pi}{k} \\
  & \quad + \frac{1}{2} (A_0^{\dagger} A_0 - A_{k+2l}^{\dagger} A_{k+2l}) \cos^2 \frac{(l-i)\pi}{k} +
     \frac{1}{2} (A_{2k+2l-i}^{\dagger} A_{2k+2l-i} - A_i^{\dagger} A_i) \cos^2 \frac{l\pi}{k} \biggr]
     \crr^{k+2l} \cii \\
  & \quad + 2b \sum_l \biggl[- (A_0^{\dagger} A_{k+2l-i} + A_i^{\dagger} A_{2l}) \sin \frac{l\pi}{k} 
     \sin \frac{(l-i)\pi}{k} + \frac{1}{2} (A_0^{\dagger} A_0 - A_{2l}^{\dagger} A_{2l}) \sin^2 \frac{(l-i)\pi}{k}\\
  & \quad + \frac{1}{2} (A_{k+2l-i}^{\dagger} A_{k+2l-i} - A_i^{\dagger} A_i) \sin^2 \frac{l\pi}{k} \biggr]
     \crr^{2l} \cii - \frac{k}{4} \sin \frac{2i\pi}{k} \sum_l \sin \frac{2l\pi}{k} (a^2 - 2ab \crr^k + b^2)
     \crr^{2l}.  
\end{split} \label{eq:H-com}
\end{equation}
It may be observed that this relation is also valid for $i=0$, leading to $[H_0, H_0] = 0$ as it should be.\par
%
%------------------------------------------------------------------------------------------------------------------------
%
On summing both sides of equation (\ref{eq:H-com}) over $i$ from 0 to $k-1$ and employing equations (\ref{eq:A-rel-1}), (\ref{eq:A-rel-2}), as well as some relations easily obtained from them, we arrive at the result $[H_0, \chh_k] = 0$, where we have taken equation (\ref{eq:H-sum}) into account. It only remains to apply $\crr^i$ and to use the invariance of $\chh_k$ under $D_{2k}$, together with equation (\ref{eq:D-H}), to infer that
\begin{equation*}
  [H_i, \chh_k] = 0, \qquad i=0, 1, \ldots, k-1,
\end{equation*}
showing that $H_i$, $i=0$, 1, \ldots,~$k-1$, are integrals of motion of $\chh_k$.\par
%
%-----------------------------------------------------------------------------------------------------------------
%
{}From them, we can form two $D_{2k}$ invariants, namely their sum proportional to $\chh_k$ and their symmetrized product
\begin{equation}
  \cyy_{2k} = (2\omega)^k \sum_p H_{p(0)} H_{p(1)} \cdots H_{p(k-1)},  \label{eq:ext-Y}
\end{equation}
where the summation runs over all $k!$ permutations of 0, 1, \ldots,~$k-1$. Projection in the $D_{2k}$ identity representation space then leads to $H_k$, on one hand, and to an integral of motion $Y_{2k}$ of the latter, on the other hand. Since $Y_{2k}$ is clearly a differential operator of order $2k$, the conjecture of \cite{tremblay09} will be proved for any odd $k \ge 3$ provided $X_k$ and $Y_{2k}$ are functionally independent.\par
%
%---------------------------------------------------------------------------------------------------------------
%
This last point can be demonstrated by showing that $[\cxx_k, \cyy_{2k}] \ne 0$ and that such a property remains true after projection. As $[\cxx_k, \cyy_{2k}]$ is a $(2k+1)$th-order operator, it is enough to prove that the coefficient of one of the highest-order terms, namely $D_{\varphi} D_r^{2k}$, does not vanish. Such a term can be obtained by multiplying the term proportional to $D_{\varphi} D_r^2$ in one the commutators $[\cxx_k, 2 \omega H_i]$ by the terms proportional to $D_r^2$ in the remaining $2 \omega H_j$'s ($j \ne i$) and summing over $i$, as well as over all permutations $p$.\par
%
%---------------------------------------------------------------------------------------------------------------
%  
On starting from the relation
\begin{equation*}
\begin{split}
  [\cxx_k, 2 \omega H_0] &= - [D_{\varphi}^2, 2 \omega H_0] \\
  &= [r^2, 2 \omega H_0] \chh_k + [(r D_r)^2, 2 \omega H_0] + 2 \Bigl[(a\crr + b) \sum_l \crr^{2l} \cii, 
         2 \omega H_0\Bigr] r D_r \\
  &\quad + 2 (a\crr + b) \sum_l \crr^{2l} \cii \,[r D_r, 2 \omega H_0] - \omega^2 [r^4, 2 \omega H_0], 
\end{split}
\end{equation*}
it can be easily shown that the coefficient of $D_{\varphi} D_r^2$ in $[\cxx_k, 2 \omega H_0]$ is equal to $- 4 \sin \varphi \cos \varphi$. It is then enough to apply $\crr^i$ to this result to prove that the term $- 4 \sin \left(\varphi + \frac{i \pi}{k}\right) \cos \left(\varphi + \frac{i \pi}{k}\right) D_{\varphi} D_r^2$ appears in $[\cxx_k, 2 \omega H_i]$. Finally, we get the term $A D_{\varphi} D_r^{2k}$ with $A = - 2^{-2k+4} k!\, k \sin k\varphi \cos k\varphi \ne 0$ in $[\cxx_k, \cyy_{2k}]$, thereby showing that $[X_k, Y_{2k}] \ne 0$.\par
%
%----------------------------------------------------------------------------------------------------------------
%
Taking into account that the truth of the conjecture of \cite{tremblay09} is already known for $k=1$ \cite{winternitz}, we conclude that it is valid for any odd $k$.\par    
%
%==============================================================================
%
\section{\boldmath The $k=3$ case}

To illustrate the approach followed in the present paper, it is worth considering the smallest allowed value of $k$, namely $k=3$. In such a case, another proof of the superintegrability of $H_k$ is indeed available through the use of other types of differential-difference operators, modified boson ones and modified boson oscillator Hamiltonians \cite{cq95}. Here we plan to provide some links between those operators and the corresponding ones constructed in sections 2, 3 and 4.\par
%
%-------------------------------------------------------------------------------------------------------------------
%
The Hamiltonian $H_3$ appears when considering the three-body Hamiltonian of the CMW problem \cite{wolfes} and eliminating the centre-of-mass motion. The latter Hamiltonian
\begin{equation*}
  H_{\rm CMW} = \sum_{i=1}^3 (- \partial_i^2 + \omega^2 x_i^2) + 2\alpha \left(\frac{1}{x_{12}^2} + 
  \frac{1}{x_{23}^2} + \frac{1}{x_{31}^2}\right) + 6\beta \left(\frac{1}{y_{12}^2} + \frac{1}{y_{23}^2} + 
  \frac{1}{y_{31}^2}\right),
\end{equation*}
where $x_i$, $i=1$, 2, 3, denote the particle coordinates, $x_{ij} = x_i - x_j$, $i \ne j$, and $y_{ij} = x_i + x_j - 2 x_k$, $i \ne j \ne k \ne i$, can indeed be separated into a centre-of-mass Hamiltonian $H_{\rm cm} = - \partial_X^2 + \omega^2 X^2$ and a relative one $H_{\rm rel}$, coinciding with $H_3$, by defining $X = (x_1 + x_2 + x_3)/\sqrt{3}$, $r \cos \varphi = x_{12}/\sqrt{2}$ and $r \sin \varphi = y_{12}/\sqrt{6}$.\par
%
%-------------------------------------------------------------------------------------------------------
%
In \cite{cq95}, $H_{\rm CMW}$ has been transformed into a $D_6$-extended Hamiltonian $\chh_{\rm CMW}$ by realizing the 12 operators of $D_6$ in terms of the particle permutation operators $K_{ij}$ and the inversion operator $I_r$ in relative coordinate space and by introducing three differential-difference operators $D_i$, $i=1$, 2, 3, defined as in equation (2.5) of \cite{cq09}. Three sets of modified boson creation and annihilation operators $a_i^{\dagger} = (\omega x_i - D_i)/\sqrt{2\omega}$, $a_i = (\omega x_i + D_i)/\sqrt{2\omega}$, $i=1$, 2, 3, have then be considered and used to build three modified oscillator Hamiltonians $h_i = \{ a_i^{\dagger}, a_i \}/2$, $i=1$, 2, 3. In terms of the latter,
\begin{equation}
  \chh_{\rm CMW} = 2\omega \sum_i h_i,  \label{eq:ext-CMW} 
\end{equation}
while the sixth-order integral of motion of $H_{\rm CMW}$ has been derived from $(2\omega)^3 \sum_i h_i^3$ by projection in the $D_6$ identity representation space.\par
%
%-----------------------------------------------------------------------------------------------------------------
%
The correspondence $K_{ij} \leftrightarrow \crr^{2k+3} \cii$ and $K_{ij} I_r \leftrightarrow \crr^{2k} \cii$, where $(ijk) = (123)$, leads to equation (2.6) of \cite{cq09}, connecting the differential-difference operators of equation (\ref{eq:D}) with the $D_i$'s. As a result, we can express the three dependent modified annihilation operators $A_i$, $i=0$, 1, 2, defined in (\ref{eq:A-i}), in terms of the three independent operators $a_i$, $i=1$, 2, 3, as $A_0 = (a_1 - a_2)/\sqrt{2}$, $A_1 = - (a_2 - a_3)/\sqrt{2}$, $A_2 = (a_3 - a_1)/\sqrt{2}$. The alternative set of operators $B_i$, $i=0$, 1, 2, is given by $B_0 = (a_1 + a_2 - 2a_3)/\sqrt{6}$, $B_1 = - (a_2 + a_3 - 2a_1)/\sqrt{6}$, $B_2 = (a_3 + a_1 - 2a_2)/\sqrt{6}$. From this, it can be easily checked that there is a single relation $A_0 - A_1 + A_2 = 0$ among the $A_i$'s (and similarly for the $B_i$'s) and that all relations in (\ref{eq:A-rel-1}) and (\ref{eq:A-rel-2}) reduce to that one for $k=3$. Furthermore, the commutation relations (\ref{eq:A-com-1}) and (\ref{eq:A-com-2}) entirely agree with those that can be derived from the above-mentioned relations between the $A_i$'s and the $a_i$'s and from the commutation relations of the latter given in equation (4.2) of \cite{cq95}.\par
%
%--------------------------------------------------------------------------------------------------------------
% 
Turning now ourselves to the modified boson oscillator Hamiltonians, we get the equations
\begin{equation}
\begin{split}
  H_0 &= \tfrac{1}{2} [h_1 + h_2 - \tfrac{1}{2} \{a_1^{\dagger}, a_2\} - \tfrac{1}{2} \{a_2^{\dagger}, a_1\}],
      \\
  H_1 &= \tfrac{1}{2} [h_2 + h_3 - \tfrac{1}{2} \{a_2^{\dagger}, a_3\} - \tfrac{1}{2} \{a_3^{\dagger},
      a_2\}],  \\
  H_2 &= \tfrac{1}{2} [h_3 + h_1 - \tfrac{1}{2} \{a_3^{\dagger}, a_1\} - \tfrac{1}{2} \{a_1^{\dagger}, a_3\}], 
\end{split}  \label{eq:H-h}
\end{equation}
where the extra anticommutators on the right-hand sides serve to compensate for the centre-of-mass degrees of freedom present in $h_1$, $h_2$, $h_3$, but absent from $H_0$, $H_1$, $H_2$. From equations (\ref{eq:H-sum}), (\ref{eq:ext-CMW}) and (\ref{eq:H-h}), it follows that
\begin{equation*}
  \chh_3 = \tfrac{2}{3} \chh_{\rm CMW} - \tfrac{1}{3} \omega \sum_{i \ne j} \{a_i^{\dagger}, a_j\},
\end{equation*}
in agreement with the decomposition
\begin{equation*}
  \chh_{\rm CMW} = H_{\rm cm} + \chh_3, \qquad H_{\rm cm} = \tfrac{1}{3} \omega \{a_1^{\dagger} +
  a_2^{\dagger} + a_3^{\dagger}, a_1 + a_2 + a_3\} = - \partial_X^2 + \omega^2 X^2.
\end{equation*}
Finally, the operator $\cyy_6$, giving rise by projection to the sixth-order integral of motion of $H_3$, could be expressed in terms of $a_i^{\dagger}$, $a_i$, $i=1$, 2, 3, through equations (\ref{eq:ext-Y}) and (\ref{eq:H-h}).\par
%
%========================================================================
%
\section{Conclusion}

In the present communication, we have provided an alternative approach to the construction of the sixth-order integral of motion of the CMW Hamiltonian, wherein the centre-of-mass motion has been eliminated from the very beginning, in contrast with the previous one based on the use of differential-difference operators related to the three particle coordinates on the line \cite{cq95}. The present construction instead employs differential-difference operators $D_r$, $D_{\varphi}$ on the relative-coordinate space. A key step in the construction has been the introduction of three dependent sets of modified boson creation and annihilation operators with simple exchange properties under the dihedral group $D_6$ connected with the CMW problem.\par
%
%---------------------------------------------------------------------------------------------------------------
%
An analogous approach, taking advantage of some recent progress made in \cite{cq09}, has allowed us to prove the superintegrability of all TTW Hamiltonians $H_k$ with odd $k \ge 3$. In such a case, the associated group being $D_{2k}$, we have been led to consider $k$ dependent sets of modified boson creation and annihilation operators and the $k$ modified oscillator Hamiltonians that can be built from them. In terms of the latter, the construction of the missing integral of motion $Y_{2k}$ has proved easy. Furthermore, $Y_{2k}$ has been shown to be functionally independent of the first integral of motion $X_k$, thereby completing the demonstration.\par
%
%-------------------------------------------------------------------------------------------------------------------------
%
Since for even $k$ values, the operator $D_{\varphi}$ assumes a different form from that considered here for odd $k$, it is not clear yet whether the present formalism can be extended to such values and lead to a proof of the remaining part of the conjecture. This is left for a future study.\par
%
%===================================================================
%
\section*{Acknowledgment}

The author would like to thank A V Turbiner for attracting her attention to the problem addressed to here and for several useful discussions.\par
%
%==============================================================================
%


\newpage
\begin{thebibliography}{99}

\bibitem{tremblay09} Tremblay F, Turbiner A V and Winternitz P 2009 {\em J.\ Phys.\ A: Math.\ Theor.} {\bf 42} 242001

\bibitem{fris} Fri\v s J, Mandrosov V, Smorodinsky Ya A, Uhlir M and Winternitz P 1965 {\em Phys.\ Lett.} {\bf 16} 354

\bibitem{winternitz} Winternitz P, Smorodinsky Ya A, Uhlir M and Fri\v s J 1967 {\em Sov.\ J.\ Nucl.\ Phys.} {\bf 4} 444

\bibitem{olsha} Olshanetsky M A and Perelomov A M 1983 {\em Phys.\ Rep.} {\bf 94} 313

\bibitem{calogero} Calogero F and Marchioro C 1974 {\em J.\ Math.\ Phys.} {\bf 15} 1425

\bibitem{wolfes} Wolfes J 1974 {\em J.\ Math.\ Phys.} {\bf 15} 1420

\bibitem{tremblay10} Tremblay F, Turbiner A V and Winternitz P 2010 {\em J.\ Phys.\ A: Math.\ Theor.} {\bf 43} 015202

\bibitem{kalnins09a} Kalnins E G, Miller W, Jr and Pogosyan G S 2009 Superintegrability and higher order constants for classical and quantum systems {\em Preprint} arXiv:0912.2278 

\bibitem{kalnins09b} Kalnins E G, Kress J M and Miller W, Jr 2009 Families of classical subgroup separable superintegrable systems {\em Preprint} arXiv:0912.3158

\bibitem{dunkl} Dunkl C F 1989 {\em Trans.\ Am.\ Math.\ Soc.} {\bf 311} 167

\bibitem{poly} Polychronakos A P 1992 {\em Phys.\ Rev.\ Lett.} {\bf 69} 703

\bibitem{brink} Brink L, Hansson T H and Vasiliev M A 1992 {\em Phys.\ Lett.} B {\bf 286} 109

\bibitem{cq95} Quesne C 1995 {\em Mod.\ Phys.\ Lett.} A {\bf 10} 1323

\bibitem{khare} Khare A and Quesne C 1998 {\em Phys.\ Lett.} A {\bf 250} 33

\bibitem{cq09} Quesne C 2009 Exchange operator formalism for an infinite family of solvable and integrable quantum systems on a plane {\em Preprint} arXiv:0910.2151, {\em Mod.\ Phys.\ Lett.} A (in press)

\end {thebibliography}
\end{document}